\documentclass[prc,twocolumn,floatfix,groupedaddress,showpacs,superscriptaddress] {revtex4}
\usepackage{graphicx}
\usepackage{dcolumn}
\usepackage{mathrsfs}
\usepackage{bm}
\usepackage{dsfont}
\usepackage{dcolumn}
\usepackage[usenames]{color}
\usepackage{amssymb}
\usepackage{amsmath} 
\usepackage{makeidx}
\usepackage{array}
\usepackage{multirow}
\usepackage{bm} 
\usepackage{soul} 
\usepackage[normalem]{ulem}
\DeclareMathAlphabet{\mathpzc}{OT1}{pzc}{m}{it} 

\usepackage{color}

\definecolor{OliveGreen}{rgb}{0,0.6,0}

\begin{document}
\title{Two-dimensional extrapolation procedure for ab initio study of nuclear size parameters and the properties of halo nucleus $^6$He }

\author{D. M. Rodkin}
\affiliation{Dukhov Research Institute for Automatics, 127055, Moscow, Russia}

\affiliation{Moscow Institute of Physics and Technology, 141701 Dolgoprudny, Moscow Region, Russia
}
\author{Yu. M. Tchuvil'sky}
\affiliation{Skobeltsyn Institute of Nuclear Physics, Lomonosov Moscow State University,
119991 Moscow, Russia}
\date{\today}

\begin{abstract}
A new two-dimensional procedure for extrapolation of the values of matter, neutron, and proton radii obtained in no-core shell model (NCSM) calculations to infinite size of its basis is proposed. A relationship between the radii is used as an additional test. Together with the JISP16 potential, which is frequently used in NCSM calculations of the radii, the Daejeon 16 potential is applied for these purposes for the first time. Halo nucleus $^6$He is the object of studies. The small spread of radii values, successful testing using relationship between the values of these three radii, and reasonable agreement between the obtained results and experimental data as well as the results of other advanced ab initio calculations demonstrate the high efficiency of the developed approach and, therefore, good prospects for its application. The performed investigations and analysis of the results of other studies indicate that the halo of $^6$He has a large size -- 0.7 $\div$ 0.8 fm.
\end{abstract}

\maketitle

The matter and charge radii are basic observables characterizing atomic nuclei. Due to a variety of experimental techniques, the charge radii of stable and long-lived nuclei have been well studied. A slightly less favorable, but also quite satisfactory situation occurs for matter radii. For nuclei far from the stability band, the study of these quantities causes serious difficulties, which are gradually overcome with the development of methods for obtaining ever more intense beams of short-lived ions. The data obtained during such measurements are of great physical importance, since they provide information about the inhomogeneity of the distribution of protons and neutrons, the presence of "skin" or "halo" in nuclei. However, such experiments are still technically difficult and their results obtained for the same isotope often differ markedly from experiment to experiment. Therefore, the problem of theoretical description the size parameters (including neutron radii not available for direct measurements) of nuclei of the discussed type seems to be very topical. At the same time, it should be stressed that simplified approaches, such as the shell model with an inert core and, even more so, schemes that consider the core without taking into account its nucleon structure, encounter difficulties in describing long-range nucleon correlations, exchange effects, etc. and, therefore, give results of limited reliability. So, the use of ab initio approaches is one of few possible ways to solve this task.

An  ab initio study of the size parameters of the lightest nucleon-stable neutron-rich $^6$He nucleus is the subject of this paper. Naturally, total binding energy of this system is also computed. The choice of this isotope is due to the fact that it is a good testing ground for the discussed studies. Indeed, first, its size characteristics have been fairly well studied experimentally. Second, the relatively small number of its constituent nucleons makes it possible to test a wide variety of high-precision theoretical methods on it. Third, this ``canonical'' example of the lightest halo nucleus turned out to be difficult to describe even within the framework of many ab initio schemes due to its large size and the great difference in neutron and charge radii (halo size $r_n-r_c$), unique for nuclei with A$\leq$10. The main motivation behind this choice is, obviously, that $^6$He nucleus is a very physically interesting (so-called Borromean) three-body  ($\alpha$ +2n) system characterized by a large distribution radius of weakly bound neutrons.

The list of works devoted to the same subject is quite large. The most important of these are presented in Refs. \cite{ai1,ai2,ai3,ai4,ai5,ai5_2,ai6,ai6_2,ai7,ai8,ai9,ai9_2,ai10,ai11} which are based on wide diversity of ab initio schemes. Various Monte Carlo approaches \cite{ai1,ai2,ai5_2,ai9_2}, hyperspherical harmonics method \cite{ai5},  no-core configuration interaction (NCCI) calculations \cite{ai8,ai9}, coupled-cluster method \cite{ai11}, etc. are used in these works. Various versions of the no-core shell model (NCSM) play the most prominent role among these schemes. The NCSM and similar in structure NCCI calculations of  matter and charge radii together with the binding energy ones are performed using realistic NN \cite{ai6} and NN + 3N   \cite{ai3,ai7,ai9_2} interactions. The bases of the usual NCCI and NCSM are exploited, as well as symmetry-adapted SU(3)-based  NCSM \cite{ai6_2}. NCSM with ($\alpha$+2n) continuum (NCSMC) \cite{ai10} also used to solve the mentioned above problem. 
 
It should be emphasized that the problem of high-precision ab initio computation of the size parameters of nuclei is qualitatively different from the problem of calculation their binding energy. The main reason for this is that the sizes of nuclei are largely determined by long-range nucleon correlations. The consequence of that is, first, that it is hard task to achieve convergence of the results of calculating the radii of nuclei on a basis that is available to modern computers. This, in turn, forces one to use procedures for extrapolating the results to large dimensions of a basis. Approaches of such a kind in the calculations of $^6$He nucleus are presented in Refs. \cite{extr1,extr2}. However, there is a second problem along this way. In contrast to the regular tendency for the binding energy to change monotonically with increasing basis size, the convergence of nuclear radii is not monotonic. Because of this, approaches that include extrapolation procedures to an infinite basis also face difficulties.

The results obtained in the above works describe the experimental data with varying degrees of success (see the analysis below). However, the approaches developed in these works are still insufficient to answer the question of whether the source of the discrepancy between the theoretical results and experiment is the model of internucleon interaction or the inaccuracy of the calculation method itself. This is true even for the canonical  $^6$He nucleus example. Therefore, the development of new methods for calculating the radii of nuclei and, in particular, for extrapolation of computed data obtained with an achievable basis cutoff parameter to infinity seems to be a vital issue.

In this paper we start from one of the most reliable and justified ab initio approaches, M-scheme of no-core shell model.  Two different versions of NN-interaction together with Coulomb proton-proton interaction have been incorporated in the model Hamiltonian. The first of these two  -- JISP16-potential -- has been built from  nucleon scattering data using the $J$-matrix inverse scattering method \cite{jm}. It has already been used to calculate the size parameters of $^6$He nucleus in Refs. \cite{ai5_2,ai6_2,ai6,ai8}, so in this paper it is exploited, in particular, to test other new elements of our approach.

An important novel of the current investigation is that the Daejeon16 potential \cite{dj16} is also applied for the NCSM computations of $^6$He nucleus radii as a model of NN-interaction. It is built using the N3LO limitation of Chiral Effective Field Theory \cite{ceft1} softened via Similarity Renormalization Group (SRG) transformation \cite{srg1}. Both JISP16 and Daejeon16 versions of internucleon interaction are ``global'' i.e. designed to calculate all kinds of characteristics of nuclei with the masses $A \leq 16$. They were tested in the framework of large-scale computations of the total binding energies, nucleon and cluster binding energies, excitation energies, radii,  moments of nuclear states and the reduced probabilities of electromagnetic transitions.  These tests demonstrated that such characteristics are, in general, reproduced well. In some cases, the quality of describing these characteristics using the Daejeon16 interaction is somewhat better than that for JISP16 one. 

The NCSM calculations were performed using the Bigstick code \cite{bigstick}.  In these computations the bases are limited by the values of the cut-off parameter ${\cal N}^*_{max} \leq 14$.  
 
The central point of this work is a new two-dimensional procedure for extrapolating the calculated values of radii to an infinite basis. The object of extrapolation is the shape of the surface of values of one of the radii over the plane (${\cal N}^*_{max}$,$\hbar \omega$) and both these coordinates are contained in the extrapolation formula.

In this way the matter, neutron and proton radii as well as the size of the neutron halo of $^6$He isotope are computed. 
A comparative analysis of the obtained results, experimental data and results of other theoretical works is given.

Let us start the description of the experimental situation and the developed approach with the terminology. As in most modern works, point-nucleon RMS radii -- parameters that characterize the distributions of all nucleons (nuclear matter) $r_m \equiv ({\bar r}^2_m)^{1/2}$, as well as neutrons $r_n$, and protons $r_p$ are considered. The last parameter is obtained from the measured charge radius $r_c$ using the expression presented in Ref. \cite{exp4}:  

\begin{equation}
r^2_{p}= r^2_c-R^2_p-(N/Z)R^2_n-3\hbar^2c^2/4(M_pc^2)-r^2_{so}.
\label{rc}
\end{equation}
Here  $R^2_p$ and $R^2_n$ are the proton and neutron mean-square charge radii, $3\hbar^2c^2/4(M_pc^2)$ is a relativistic
Darwin-Foldy correction, and $r^2_{so}$ is a spin-orbit nuclear charge-density correction. The following values are usually chosen: $R^2_n= - 0.1161$ fm$^2$,  $3\hbar^2c^2/4(M_pc^2)$ = 0.033 fm$^2$, $r^2_{so}$ = 0.08 fm$^2$. The last estimate is valid for $^6$He only. 
For $R_p$ the Particle Data Group \cite{prot1} value is 0.877(7) fm. It is this quantity that is used when processing the measurement results to obtain point-proton nuclear radii. However, this value has been also precisely determined from the spectroscopy of muonic hydrogen \cite{prot2}. It turned out to be 0.84184(67) fm. Therefore, using these two values for $R_p$ two different values of  $r_p$ can be obtained. Matter radius $r_m$ is deduced directly from differential cross sections of elastic proton scattering at high momentum transfer using Glauber multiple-scattering theory. Radius $r_n$ is unavailable for measurements and therefore is calculated using the expression
\begin{equation}
Ar^2_m= Zr^2_p+Nr^2_n.
\label{rn}
\end{equation}

As in other papers when studying the properties of systems with a neutron halo or skin, a measure of their thickness -- the difference of point-nucleon sizes $r_h=r_n - r_p$ is used by us. In NCSM computations we use the universal approach, which makes it possible to calculate all three size parameters. After calculating the binding energy and wave function, we proceed for calculating matter, neutron and proton radii for point nucleons. The basic expressions of the formalism intended to describe these quantities are the following. In the shell model, the squared radius of a corresponding system is defined as
 
\begin{equation*}
 r^2_{m(n,p)} =  (1/N_A) \sum_{i} (\vec{r}_{m(n,p),i} - \vec{r}_{cm})^2,
\end{equation*} 
where $\vec{r}_{cm} = (1/N_A) \sum_{i} \vec{r}_{m,i}$
Mean square radius takes the form 

\begin{equation}
\begin{aligned}
{\bar r}^2_{m(n,p)} = - \frac{4}{N_A\cdot N_{A(N,Z)}} \langle \Psi_{A} | \sum_{i <j} \vec{r}_{m(n,p,i)} \vec{r}_{m,j} | \Psi_A{} \rangle\\ +\langle \Psi _{A}| r_{cm}^2 | \Psi _{A} \rangle + \frac{N_A - 2}{N_A\cdot N_{A(N,Z)}} \langle \Psi_{A} | \sum_{i} r^{2}_{m(n,p),i} | \Psi_A{} \rangle. 
\end{aligned}
\end{equation}
Here $N_{A(N,Z)}$ denotes the number of nucleons $A$ (or neutrons $N$, or protons $Z$) in the system,

\begin{equation}
\langle \Psi _{A}| r_{cm}^2 | \Psi _{A} \rangle = \frac{3 (\hbar c)^2}{2mc^2 \hbar \omega N_A},
\end{equation}

\begin{equation}
\begin{aligned}
\langle \Psi _{A}| \sum_{i} \vec{r_i}^2 | \Psi _{A} \rangle = \frac{1}{\sqrt{2J+1}} \\
\sum_{k_a, k_b} OBTD(k_a, k_b, \lambda = 0) \langle k_a || r^2 || k_b \rangle,
\end{aligned}
\end{equation}

and
\begin{equation}
\begin{aligned}
\langle \Psi_{A} | \sum_{i < j} \vec{r}_{i} \vec{r}_{j} | \Psi_A{} \rangle = \frac{1}{\sqrt{2J +1}}
\displaystyle \sum_{k_a \le k_b, k_c \le k_d, J_0}\\
 \langle k_a k_b J_0 || \vec{r}_{1} \vec{r}_{2} || k_c k_d J_0 \rangle \cdot TBTD(k_a, k_b, k_c, k_d, J_0).
\end{aligned}
\end{equation}

The one-body and two-body transition densities (OBTD) and (TBTD) included in these formulas are expressed in terms of the matrix elements of the products of fermion second quantization operators: 
\begin{equation}
OBTD(k_a, k_b, \lambda = 0) = \langle \Psi_{A} || [a^{+}_{k_a} \otimes \tilde{a}_{k_b}]^{\lambda = 0} || \Psi_{A} \rangle.
\end{equation}

\begin{equation}
\begin{aligned}
TBTD(k_a, k_b, k_c, k_d, J_0) = \langle \Psi_{A} ||[ [a^{+}_{k_a} \otimes a^{+}_{k_b}]_{J_{0}} \otimes \\
[ \tilde{a}_{k_c} \otimes \tilde{a}_{k_d}]_{J_{0}} ]^{\lambda = 0} || \Psi_{A} \rangle.
\end{aligned}
\end{equation}
where $ \otimes$ is the sign of a tensor product of rank $J_0$ or $\lambda$.

The NCSM calculations of the  radii are supplemented by the two-dimensional extrapolation procedure proposed by us. The results of calculations of these quantities for different values of $\hbar \omega$ and  ${\cal N}^*_{max}$ are used as input data. To understand the reasoning behind this idea consider the surface formed by the values of any of the investigated radii over the plane (${\cal N}^*_{max}$,$\hbar \omega$). It is clear that for large values of $N^{*}_{max}$ this surface is a horizontal plane in a fairly wide range of values of $\hbar \omega$. As ${\cal N}^*_{max}$ decreases, the values of the radius increase for small $\hbar \omega$ and decrease for large $\hbar \omega$, which was noted in several previous works, in particular in Refs. \cite{ai6,ai8} and stressed by illustrations of the latter one. The wide horizontal part of the surface narrows and, in the range of  maximal ${\cal N}^{*}_{max}$ values achievable for computer calculations, degenerates into an (approximately) horizontal line in some domain of values ${\cal N}^{*}_{max}$. The values of the radius on this line, it was proposed to consider as a value of the corresponding size parameter. It is so-called ``crossover prescription''. The value of $\hbar \omega$ at which the value of this parameter stabilizes is called crossover point. A detailed description of this concept can be found in Ref. \cite{ai8}. The results of our calculations with the JISP16 potential, naturally, give the same pattern as the results of the papers just cited, and, importantly, an analogous picture is observed for the Daejeon16 potential. The geometric image of the surface as a whole in this case is a tape twisted around this line as an axis. So, the idea is to determine trends in surface properties with increasing ${\cal N}^*_{max}$ using an extrapolation procedure basing on two-dimensional data with limited values of this parameter. We called this procedure twisted tape extrapolation (TTE). Within the framework of the TTE procedure, the following formula is proposed:

\begin{equation}
\begin{aligned}
r^2_{m(n,p)}({\cal N}^*_{max},\hbar \omega ) = r^2_{\infty,m(n,p)} + \\
P_k(\hbar \omega )\exp ( - \alpha \sqrt{{\cal N}^*_{max}}),
\label{extr}
\end{aligned}
\end{equation}
where $P_k(x)$ -- a polynomial of degree $k$ whose coefficients are fitting parameters, $r^2_{\infty,m(n,p)}$ is the extrapolation result -- squared radius for infinite oscillator basis and $ r^2_{m(n,p)}({\cal N}^*_{max},\hbar \omega )$ are theoretically obtained results for squared radii. Using in the expression (\ref{extr}) an exponential dependence with an index proportional to $[{\cal N}^*_{max}]^{1/2}$, we proceed according to the conclusions of Ref. \cite{extr2}. Factor of proportionality $\alpha$ is also a fitting parameter.

It should be noted that the interpolation procedure to determine the ranges of possible values of the radii $r_p$ and $r_m$ using two-dimensional  (${\cal N}^*_{max}$,$\hbar \omega$) input data was presented in Ref. \cite{ai8}. Namely, the radii as functions of $\hbar \omega$  at fixed ${\cal N}^*_{max}$ values from the domain of approximate stability are computed by cubic one-dimensional interpolation of the calculated data points at different $\hbar \omega$. A more complicated  two-dimensional-basing  extrapolation procedure which includes the chi-square fit to the number of points at which the values of the point-proton radius were calculated for various ${\cal N}^*_{max}$ and $\hbar \omega$ is presented in Ref. \cite{extr_meth}. The object of the study was the $^6$Li nucleus. Some part of the calculations was carried out with the JISP16 potential. The weight of each point was determined by the difference in the values of the radius in neighboring ${\cal N}^*_{max}$ points. 

The main novel of our procedure is that the interdependence of  $r^2_{\infty}$ and $\hbar \omega$ is directly included in the extrapolation formula (\ref{extr}). Another difference of our extrapolation procedure is a choice of the range of $\hbar \omega$ values which is  fundamentally different from the one that is preferred in Ref. \cite{extr_meth}. It seems to us preferable to search in the vicinity to crossover point as it is done in work \cite{ai8}. At the same time, in our work, we used the same method for determining the weight of each point, as proposed in Ref. \cite{extr_meth}. For implementing the extrapolation procedure we have chosen chi-square fit method realized in TMinuit minimization package which is included in open-source ROOT CERN data analysis framework. 

In this paper we present the results of the first test of TTE which is a rather general and, in our opinion, promising approach for studying observables that reflect long-range internuclear correlations in various nuclei. Therefore, here we limit ourselves to the simplest linearized version of the $\hbar \omega$-dependence of the surface shape, in which $\hbar \omega$-dependence is given by a first-order polynomial $P_1(\hbar \omega )$,  and a relatively narrow mesh of calculated data. We have included in it only three values of ${\cal N}^*_{max}$ in which the crossover conditions take place: 10, 12, and 14.  It is clear from our results and known from  Refs. \cite{ai6,ai8} that as $N^{*}_{max}$ decreases, the values of the radii increase for small $\hbar \omega$ significantly faster than decrease for large $\hbar \omega$. To take that into account we have used non-uniform mesh with the points at: $\hbar \omega$ = 7.5; 8; 9; 10; 11; 12.5; 15; 17.5 MeV -- for the JISP16 potential and at $\hbar \omega$ = 6; 7.5; 8; 9; 10; 11; 12.5; 15; 17.5 MeV -- for the  Daejeon16 potential. This is partly justified by the results obtained. The prospects for TTE lie, obviously, in the field of using polynomials of a higher degree $\hbar \omega$, a denser and wider mesh. This will require a large amount of computer time. 

The results of NCSM calculations in which the JISP16 potential has been used are the following. Positions of the crossover points, values of the total binding energy (TBE) and the radii in these points are presented in Tab. I. It can be seen that these positions for the matter (as well as neutron) and point-proton radii differ noticeably. In view of this fact, the measure of violation of the relation (\ref {rn}) is important to estimate the reliability of these results. This measure is chosen to be $\Delta=1-[(Zr^2_p+Nr^2_n)/Ar^2_m]^{1/2}$. We call it violation factor. In the discussed case it is equal to 0.67 \%. 

TBEs $E_{tot}$ of $^6$He nucleus in these points are strongly underestimated compare to the experimental one. From a formalistic point of view, the system appears to be unbound with respect to the emission of two neutrons. When computing TBE, the optimal value of $\hbar \omega$ is 17.5 MeV with $E_{tot}$ = 28.47 MeV at this point, the system is bound, but the binding energy of neutrons is very small.

\begin{center}
\begin{table}
\caption{Approximate positions of the crossover points (MeV) and the values of: the total binding energy (MeV), the matter, neutron,  and point-proton radii of $^6$He nucleus (fm) at this point at ${\cal N}^*_{max}$ = 14 for the JISP16 interaction.}
\begin{tabular*}{0.22\textwidth}{ c c c c }
\hline\hline\noalign{\smallskip}
System &nucl. & neutr. & prot.\\
\hline\noalign{\smallskip}
 $\hbar \omega_{co}$& 10 &10 & 12.5 \\
$E_{tot}$ &26.41  &26.41 &27.86\\
$r$ &2.334 &2.550 &1.797 \\
\hline\noalign{\smallskip}
\end{tabular*}

\end{table}
\end{center}
 
In order to estimate stability of the results of extrapolation of size parameters, the domain of $\hbar \omega$ values has been varied. Namely, the procedure has been performed throughout the above presented range from 7.5 to 17.5 MeV, which is denoted as J0, as well as  throughout the narrowed ranges 7.5 $\div$ 15 MeV (J1), 8 $\div$ 15 MeV (J2), and 9 $\div$ 17.5 MeV (J3). The final (optimal) result is the one that corresponds to the smallest value of $\chi^2$.

\begin{figure}[htp]
\includegraphics[scale=0.42]{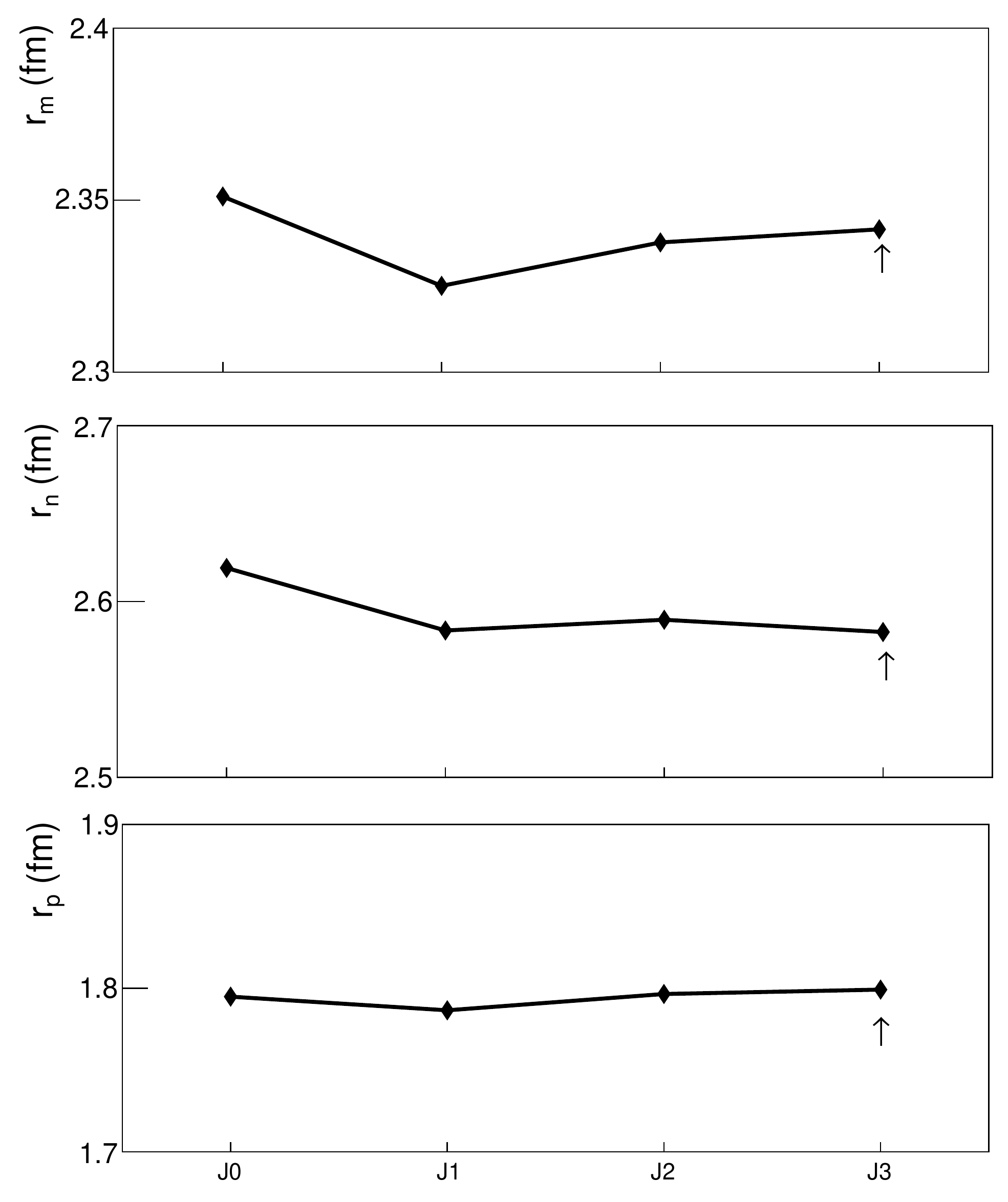} 
\caption{Matter (upper panel) neutron (middle panel) and point-proton (lower panel) radii of $^6$He nucleus for different $\hbar \omega$ extrapolation domains in case of the use of the JISP16 potential.}\label{fig1}
\end{figure}

Extrapolation results for each of these ranges in case of the use of the JISP16 potential are presented in Fig.  \ref{fig1}. The optimal values are marked with an arrow. The figure demonstrates high stability of the obtained radii. The final results for each of them and RMS deviations of other ones from the optimal are: $r_m = 2.342(7)$ fm, $r_n = 2.582(3)$ fm, and $r_p = 1.799(6)$ fm. The extrapolated quantities satisfy relation (\ref{rn}) much better than the ones estimated within the crossover prescription, violation factor $\Delta$ is equal to --0.36 \% for them. This property and, especially, the small values of the standard deviation are, in our opinion, evidence of the high efficiency of the proposed extrapolation method, even in its minimal version. 

The proposed procedure somewhat changes the value of $r_n$, and practically does not change the values of other size parameters obtained within the framework of stable crossover points prescription. The results obtained in our calculations, both estimated within the crossover prescription (see Tab. I) and extrapolated, are in rather good agreement with the results of Ref. \cite{ai8}, also obtained using the JISP16 interaction in different ways, but with up to 16 ${\cal N}^*_{max}$ values and based on the crossover prescription.  The values $r_p =1.799 \div 1.810$ fm and $r_m =2.314 \div 2.327$ fm are presented in this paper.

\begin{figure}[htp]
\includegraphics[scale=0.42]{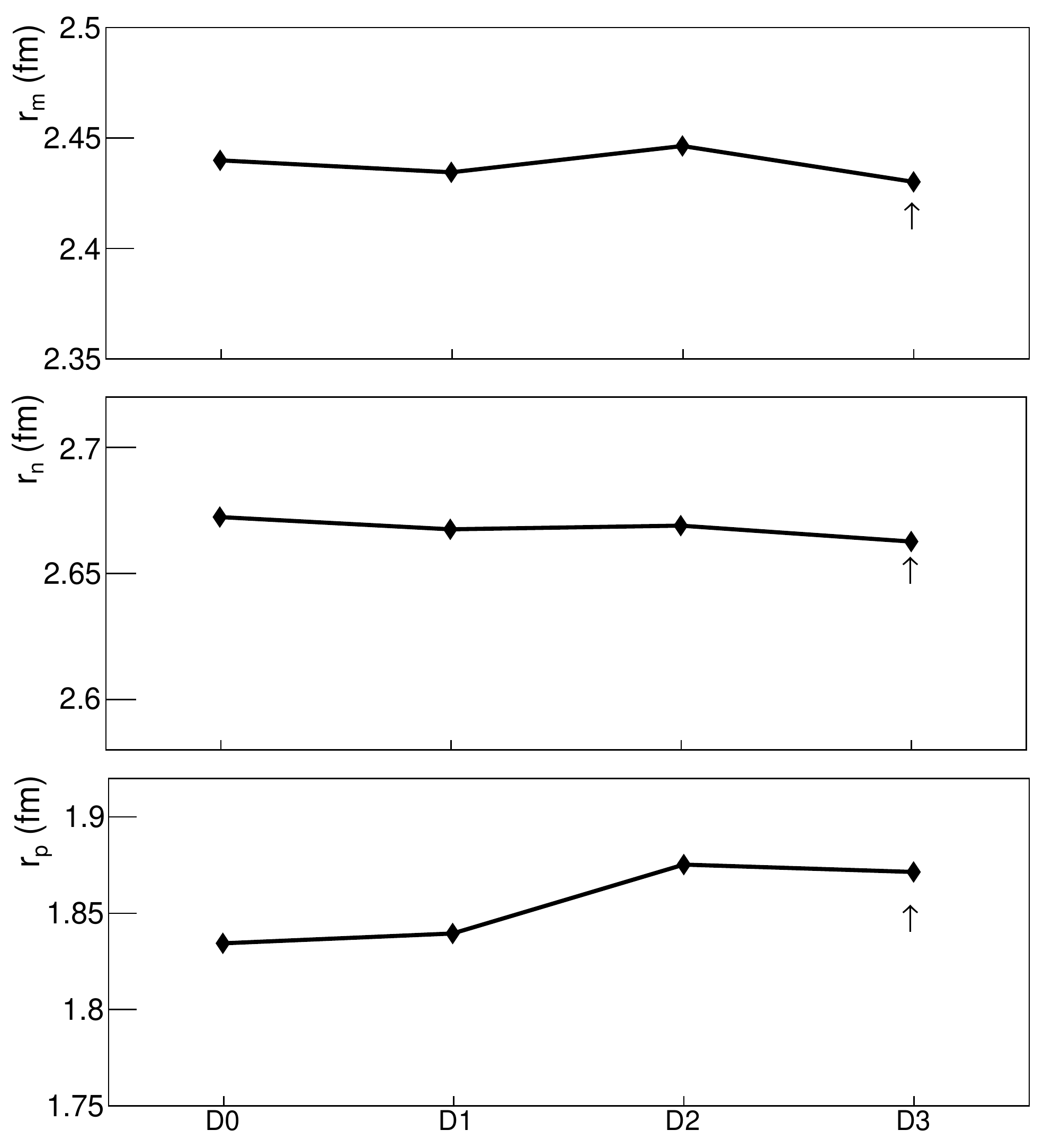} 
\caption{The same as in Fig. \ref{fig1} in case of the use of the Daejeon16 potential.}\label{fig2}
\end{figure}

Analogous computation results in which the Daejeon16 potential has been exploited are presented in Tab. II and Fig. \ref{fig2}. As for the JISP16 potential, in this case one can observe a difference in the position of the crossover points. The violation factor is significantly better being equal to --0.26 \%. For $\hbar \omega$ = 8 MeV the binding energy of two neutrons is approximately equal to 0.5 MeV, and for $\hbar \omega$ = 10 MeV it is close to that measured experimentally. At $\hbar \omega$ = 12.5 MeV, the value of $E_{tot}$ reaches a maximum value of 29.31 MeV.

\begin{center}
\begin{table}
\caption{The same as in Tab. I  for the Daejeon16 interaction.}
\begin{tabular*}{0.22\textwidth}{ c c c c }
\hline\hline\noalign{\smallskip}
System &nucl. & neutr. & prot.\\
\hline\noalign{\smallskip}
 $\hbar \omega_{co}$&8 &8 & 10 \\
$E_{tot}$ &28.83 &28.83 &29.26\\
$r$ &2.416 &2.630 &1.880 \\
\hline\noalign{\smallskip}
\end{tabular*}

\end{table}
\end{center}

The extrapolation procedure has been performed throughout the total range 6 $\div$ 17.5 MeV (D0) and the narrowed ranges 6 $\div$ 15 MeV (D1), 7.5 $\div$ 15 MeV (D2), and 7.5 $\div$ 12.5 MeV (D3). The optimal values have been chosen as described above. The obtained results are also stable. These values of the radii and the RMS-deviations are: $r_m = 2.430(6)$ fm, $r_n = 2.663(3)$ fm, and $r_p = 1.871(16)$ fm. In this case the violation factor $\Delta$ is very small, namely, equal to 0.09\%.

The experimental data necessary for comparative analysis are presented in Tab. III. The neutron radius given together with the matter radius was calculated by the authors of the experiments using their own values of $r_m$, the values of point-proton radius from \cite{exp4}$^a$ or \cite{exp6}, and relation (\ref {rn}). With reference to the table, it can be seen that the point-proton radius of $^6$He nucleus  is extracted with a high degree of accuracy and the results of different research groups are in good agreement. It looks a little strange, but against the background of this good agreement between these data, two different versions of the proton radius $R_p$ introduce noticeable duality. The data concerning the matter radius also agree well. The error bars, however, are substantially higher in these measurements. The datum of Ref. \cite{exp7} somewhat fall out of the systematics, which has a particularly noticeable effect on the size of the halo.

A comparison of the radii obtained for the two versions of the internucleon interaction with each other and with the experimental data leads to the following conclusions. The radii calculated using the Daejeon16 and JISP16 potentials differ significantly and, considering RMS deviations, reliably. Daejeon16-based calculations  result in larger values of the size parameters. The most likely reason is that the Daejeon16 interaction is softer than  JISP16. A very unexpected result is almost complete coincidence of the size of the neutron halo $r_h$. Indeed, this value is equal to 0.792 and 0.783 fm in the discussed cases.  In our opinion, this is not an artifact, but a real physical result, and an important one at that. This coincidence also holds for the results obtained via crossover prescription, $r_h$ is equal to 0.750 and 0.733 fm in this approach.

\begin{center}
\begin{table}
\caption{Experimental values of matter, neutron, and  point-proton radii of $^6$He nucleus (fm), obtained using the  radius  of proton $R_p$ from: \cite{prot1} $^a$  and \cite{prot2} $^b$.}
\begin{tabular*}{0.35\textwidth}{ c c c c c }
\hline\hline\noalign{\smallskip}
&\cite{exp1} & \cite{exp2}& \cite{exp7}& \cite{exp8}\\
\hline\noalign{\smallskip}
$r_m$ & 2.33(4) & 2.30(7) & 2.44(7) & 2.29(6) \\
$r_n$  &2.51(6) & 2.47(10)& 2.66(10)	& 2.45(9) \\
\hline\hline\noalign{\smallskip}
 &\cite{exp3,exp5} & \cite{exp4}$^a$&\cite{exp4}$^b$&\cite{exp6} \\
 \hline\noalign{\smallskip}
$r_p$ & 1.925(12) & 1.938(23)  & 1,953(22) &1.934(9)  \\

\hline\noalign{\smallskip}
\end{tabular*}

\end{table}
\end{center}

For the Daejeon16 potential, the point-proton radius is slightly, but reliably less than the experimental value. For the JISP16 potential this underestimation is not small. Other theoretical papers also give a significant underestimation of the parameter under discussion: \cite{ai7} -- 1.82 fm,  \cite{ai5} -- 1.78 fm and 1.82 fm, \cite{ai9_2} -- 1.74 fm, 1.81 fm, and 1.84 fm (depending on the choice of interaction version). In the last example the values presented in the original paper for the charge radius $r_c$ have been recalculated by us according to formula (\ref{rc}) with $R_p$ value given from Ref. \cite{prot1}. These works also confirm the trend towards an increase in $r_p$ when using softer interaction options. The exception is part of the data presented in Ref. \cite{ai11}. As a result of NCSM calculations in the basis limited by ${\cal N}^*_{max}$=12 using potential from Ref. \cite{ceft1} softened via the SRG procedure with parameter $\lambda_{SRG}$ =1.5 fm$^{-1}$  (for  $\hbar \omega$ = 14 MeV basis) and 2.0 fm $^{-1}$  (for  $\hbar \omega$ = 20 MeV basis) values 1.79 and 1.74 fm were obtained for the  point-proton radius, respectively. At the same time in the NCSMC calculations carried out for the same input data, much larger values of the radius 1.85 and 1.87 fm were obtained, and the trend of their change turned out to be opposite. In the most advanced version of NCSMC, i.e. for the biggest model space the value 1.90(2) fm which is close to the experimental one was achieved.

The presented above values of the matter radius obtained in our calculations for both versions of the interaction lie in the range of values presented by different experimental groups. One can see good agreement between the values obtained in the calculations using JISP16 potential and the data from Refs. \cite{exp1,exp2,exp8}, as well as the value obtained in the calculations using the Daejeon16 potential and the datum from Ref. \cite{exp7}. The trend for a size parameter value to increase when using softer versions of the interaction remains valid for the matter radius. This parameter was also studied in work \cite{ai11}. The matter radius obtained in the NCSM calculations with the above presented input turned out to be equal to 2.25 and 2.15 fm for $\lambda_{SRG}$ =1.5 fm$^{-1}$ and 2.0 fm$^{-1}$, respectively. At the same time the  NCSMC calculations performed in this work gave results of 2.37 and 2.41 fm, respectively, i.e. they sharply increase the values of the matter radius compared to NCSM calculations and reverse their dependence on the hardness of interaction. The computations within the most advanced version of NCSMC resulted in value $r_m = 2.46(2)$ fm.

It is important to ask which of the values of the matter radius obtained in experiments is confirmed in ab initio calculations: the larger one, 2.44(7) fm obtained in work \cite{exp7}, or the smaller ones, coinciding with good accuracy, (mean value is approximately equal to 2.31(6) fm) presented in Refs. \cite{exp1,exp2,exp8}. For several reasons, we, evidently, give preference the results obtained with the use of the Daejeon16 interaction.  Compared to JISP16 one, this Hamiltonian is newer, provides faster convergence of TBEs of various nuclei. A lot of other nuclear observables are more reproductive by it. What about $^6$He example it results in correct binding energies of two neutrons in this isotope and yields better value of the point-proton radius. The extrapolated values of radii obtained with the use of it better satisfy relation (\ref{rn}). Therefore, our study can be considered as a theoretical confirmation of the results of measurements of the matter radius presented in Ref. \cite{exp7}. The results of NCSMC calculations presented in Ref. \cite{ai11} are also in good agreement with these two values of the matter radius. 

The same question concerning the size of neutron halo is, perhaps, even more intriguing. The measurements presented in Refs. \cite{exp1,exp2,exp8} resulted in well-consistent small values of $r_h$: 0.57, 0.53, and 0.51 fm. The result of paper \cite{exp7} is much greater -- $r_h$ = 0.72 fm. The last value is in reasonable agreement with the results of calculations using the Daejeon16 potential, not only within the framework of the proposed extrapolation procedure, but also within the framework of the crossover prescription. It is noteworthy that this is also true for calculations using the JISP16 potential. An additional confirmation the discussed result may be found in the framework of the analysis of the results presented in  Ref. \cite{ai11}. The value of neutron radius $r_n$= 2.70 fm obtained by us on the basis of the data of this work using  formula (\ref{rn}) is, evidently, in good agreement with the data from Ref. \cite{exp7} and the results of our calculations. So, the computations performed in this work for two versions of the internucleon interaction, the results of advanced NCSMC computations from paper \cite{ai11} and the experimental data presented in Ref. \cite{exp7} give well-consistent large values of the size of the neutron halo of $^6$He nucleus $r_h$: 0.792, 0.783, 0.80, and 0.72 fm, i.e. about one and a half times larger than those presented in Refs. \cite{exp1,exp2,exp8}.

In conclusion let us list the basic points of the performed investigations.

\noindent
I. A new two-dimensional procedure for extrapolation of the values of matter, neutron, and proton radii obtained in no-core shell model calculations, using various harmonic oscillator bases characterized by different parameters of ${\cal N}^*_{max}$ and $\hbar \omega$, to infinite basis size is proposed. The extrapolation formula contains both these parameters together.

\noindent
II. In order to estimate stability of the results of extrapolation of size parameters, the domain of $\hbar \omega$ values has been varied. A relationship between the values of these three radii is used as an additional test.
\noindent

\noindent 
III. The JISP16 and Daejeon16 internucleon interactions are used in NCSM computations of halo nucleus $^6$He. The latter one is involved to the calculations of radii for the first time.

\noindent
IV. The small values of the RMS deviations together with successful testing using a relationship between the values of these three radii, and reasonable agreement between the obtained results and experimental data as well as the results of other advanced ab initio calculations demonstrate the high efficiency of the developed approach. This merits of it allows one, probably, in many cases to compare the quality of the description of the size parameters of nuclei by different Hamiltonians. 

\noindent
V. The results of computations of the size of $^6$He nucleus halo turns out to be the very stable and almost the same for the JISP16 and Daejeon16 potentials. The performed investigations and analysis of the results of other ab initio studies indicate that the halo of $^6$He has a large size -- 0.7 $\div$ 0.8 fm. These results confirm the material radius measurement datum presented in Ref.  \cite{exp7}.

\noindent
VI. Based on the features of the geometry of surfaces of radii, the method as a whole was called twisted tape extrapolation. In our opinion, it looks rather general and promising approach for studying observables that reflect long-range internuclear correlations in various nuclei. 

The study is supported by a grant from the Russian Science Foundation No. 22-22-00096, https://rscf.ru/en/project/22-22-00096/.
We are grateful to A. M. Shirokov for providing the Daejeon16 NN-potential matrixes, as well as  to C. W. Johnson for supporting our efforts to introduce code Bigstick for NCSM calculations.

\end{document}